\documentclass[11pt]{article}
\def\ds{\displaystyle}
\def\ss{\scriptstyle}

\usepackage{amsmath}
\usepackage{amsfonts}
\usepackage{amssymb}\usepackage{amsmath}
\usepackage{amsfonts}
\usepackage{amssymb}
\begin{document}
\pagestyle{headings}
\renewcommand{\thefootnote}{\alph{footnote}}

\title{Classical limits of unconstrained QFT}
\author{Glenn Eric Johnson\\Oak Hill, VA.\\E-mail: glenn.e.johnson@gmail.com}
\maketitle

{\bf Abstract:} In nonrelativistic limits for states labeled by minimum packets with constrained spatial spreads and over a short term, states of unconstrained quantum field theories evolve on trajectories described by Newton's equations for the $1/r^2$ force. These states include bound solutions in the attractive force case.


%= = = = = = = = = = = = = = = = = = = = = = =
\section{Introduction}
%= = = = = = = = = = = = = = = = = = = = = = =

Unconstrained quantum field theories (UQFT) [\ref{gej05}] generalize the original Wightman functional analysis development of relativistic QFT [\ref{wight},\ref{borchers}]. UQFT have explicit realizations that exhibit interaction but the quantum fields are not Hermitian Hilbert space operators. Since canonical quantization does not apply for UQFT, alternative classical limits of the relativistic quantum dynamics are of interest.

This note is a demonstration that the constructed relativistic, local quantum field theories exhibit classical limits. The UQFT constructions include states that are interpretable as scattered and bound classical particles that follow trajectories that are solutions to Newton's equation for $1/r^2$ forces. The demonstration is for UQFT with a single, neutral, Lorentz scalar field. These nonrelativistic, classical limits do not result from conjecture that quantum dynamics is described by canonical quantization of classical dynamics, that is, do not result from a correspondence of classical dynamical quantitites with Hilbert space operators. Indeed, the UQFT Hamiltonian would be a ``free field'' Hamiltonian in a canonical quantization development. Here, the correspondence is that the support of particular states follow classical particle trajectories. The abandonment of canonical quantization enables the consistent description of relativistic QFT that coincide with nonrelativistic, classical mechanics in appropriate limits. Classical particles are associated with spatial concentrations in the support of elements of the relativistic QFT. The observables of the relativistic QFT include the arguments, in particular, spacetime and energy-momentum, of the functions that label the states [\ref{johnson}].

The demonstration is developed in two steps:\begin{itemize}\item[--] In Section \ref{sec-minpak}, it is demonstrated that selected minimum packet functions approximate solutions to a Schr\"{o}dinger equation when time is the Newtonian time parameter $\tau$ describing classical trajectories for a $1/r^2$ force.
\item[--] In Section \ref{sec-schro} and in the same nonrelativistic, classical limit as Section \ref{sec-minpak}, it is demonstrated that the peaked spacetime support of UQFT states labeled by the minimum packet functions evolve along classical trajectories in the short term.\end{itemize}No constraints in addition to the general form of the constructed Wightman-functional result from the short term approximation of nonrelativistic, classical dynamics-derived descriptions. Additional constraints on the form of the UQFT Wightman-functional do result from equivalence of scattering cross sections. In [\ref{feymns}], it was demonstrated that UQFT scattering cross sections approximate the first Born approximation of nonrelativistic, canonical quantization-derived scattering cross sections for a mollified $1/r^2$ force between distinguishable particles. Scattering cross sections result from long term, plane wave limits of transition amplitudes. The $1/r^2$ force appears naturally as the long range interaction of particle pairs in the constructed UQFT. In this sense, weak gravitational forces appear naturally in UQFT that exhibit interaction.

The constructions of UQFT establish that: there is a correspondence of classical mechanics and the constructed generalizations of Wightman-functionals; bound states appear for attractive potentials; and nonrelativistic limits of UQFT cross sections agree with the results of canonical quantization. The realizable UQFT generalizations of Wightman-functionals are compatible with nonrelativistic classical limits and ordinary quantum mechanics for the $1/r$ potentials.

After a digression to describe notation in Section \ref{sec-notation}, the demonstrations for a single, neutral, Lorentz scalar field are presented in Sections \ref{sec-minpak} and \ref{sec-schro}. To further fix notation and provide context, descriptions of Jacobi coordinates, solution of Newton's equation for the $1/r^2$ force, and the Wightman-functional and scattering amplitudes for single, neutral, Lorentz scalar UQFT are included in Appendices \ref{app-jacobi}-\ref{app-uqft}.

%= = = = = = = = = = = = = = = = = = = = = = =
\section{Classical particle limits of UQFT}
%= = = = = = = = = = = = = = = = = = = = = = =
The classical limits of the constructed UQFT are studied using states labeled by minimum packet functions with constrained spatial packet spread. In limits for the selected packet spreads, a representative of the arguments of the peaked support of the states can be interpreted as a classical particle evolving along a trajectory. The particle trajectories are described by two particle solutions of Newton's equations for the conservative, central, $1/r$ potential. Nonrelativistic classical mechanics provides reference frame dependent, short term approximations to the relativistic UQFT. The free field contributions to the Wightman functions include factors $\delta(p_{\mathit{in}}-p_{\mathit{out}})$ for pairings of incoming and outgoing momenta and these factors result in straight line classical trajectories, trajectories that exhibit no interaction.

First, a digression to establish notation.

\subsection{Notation}\label{sec-notation}

The development generally follows Borchers' description of scalar QFT [\ref{borchers}]. The properties of the Wightman-functional $\underline{W}=(W_0,W_1,W_2((x)_2)\ldots)$ defines a UQFT for terminating sequences of functions $\underline{f}=(f_0,f_1(x_1),f_2((x)_2),\ldots)\in {\cal A}$ with $f_0\in {\bf C}$ and $\tilde{f}_n((\pm \omega,{\bf p})_n)\in {\cal S}({\bf R}^{3n})$, the Schwartz tempered functions [\ref{gel2}], and the component generalized functions have Fourier transforms of the form\[\tilde{W}_n((p)_n=\sum_{(s)_n} T_{(s)_n}(({\bf p})_n)\;\prod_{k=1}^n \delta(E_k\!-\!s_k \omega_k)\]with the $T_{(s)_n}(({\bf p})_n)\in {\cal S}'({\bf R}^{3n})$ and the summation is over all $2^n$ possibilities for the signs $s_k=\pm 1$.\[\omega_j^2:=\left(\frac{mc}{\hbar}\right)^2+{\bf p}_j^2\]with a mass $m>0$ and $E_j^2=\omega_j^2$ describe mass shells in ${\bf R}^4$. $E_j=\omega_j$ is the positive mass shell and $E_j=-\omega_j$ is the negative mass shell. $\underline{W}$ satisfies Poincar\'{e} covariance and microcausality for sequences from ${\cal A}$ and provides a semi-norm for sequences from ${\cal B}\subset {\cal A}$, functions with Fourier transforms supported only on positive energies. $\underline{f}\in {\cal B}$ when\begin{equation}\label{B-defn}\tilde{f}_n((p)_n) = \prod_k (E_k+\omega_k) \tilde{\varphi}_n((p)_n)\end{equation}with $\varphi_n((x)_n)\in {\cal A}$ [\ref{gej05}]. Fields are identified as multiplication in the algebra of function sequences ${\cal A}$ and for free fields, the field is the established, Hermitian Hilbert space operator. For the constructed UQFT, the Hamiltonian is the result of\begin{equation}\label{hamil}U(t)\tilde{f}_n((p)_n = \prod_{k=1}^n e^{-i\omega_k t} \,\tilde{f}_n((p)_n,\end{equation}designated the free field Hamiltonian in canonical quantization-based QFT.

Spacetime coordinates in four dimensions are designated $x:=t,{\bf x}$ with ${\bf x}:=x,y,z$ and $x$ is defined by context. Energy-momentum vectors are $p:=E,{\bf p}$. $x,p \in {\bf R}^4$, ${\bf x},{\bf p}\in {\bf R}^3$ and $x,p$ are Lorentz vectors. $x^2:=t^2-\|{\bf x}\|^2$, $p^2:=E^2-\|{\bf p}\|^2$ and $px:=Et-{\bf p}\!\cdot\!{\bf x}$ use the Minkowski signature, ${\bf p}\!\cdot\!{\bf x}$ is the Euclidean dot product and $\|{\bf x}\|^2$ is the square of the Euclidean length $\|{\bf x}\|$ in ${\bf R}^3$. In this note, the units of spacetime coordinates are length, and the units of the energy-momentum coordinates are inverse length. Conversion of $t$ to units of time is then ``time'' = $t/c$. ``momentum'' = $\hbar p$ and energies are ``energy'' = $\hbar c E= \hbar c \sqrt{(mc/\hbar)^2+{\bf p}^2}$. To ease comparison with the Schr\"{o}dinger equation,  in this note mass $m$ is in kilograms rather than inverse length. Multiple arguments include an identification index. Ascending or descending sequences of multiple arguments are denoted $(x)_{j,k}:= x_j,x_{j+1},\ldots x_k$ in the ascending case and $(x)_n:=(x)_{1,n}$. $\tilde{f}_n((p)_n)$ denotes the Fourier transform of $f_n((x)_n)$. The definition of Fourier transform adopted here is the evident multiple argument extension of\begin{equation}\label{fourier}\tilde{f}(p):= \int \frac{dx}{(2\pi)^2}\; e^{-ipx} f(x)\end{equation}and $\tilde{T}(\tilde{f}):=T(f)$. Summation notation is used for generalized functions,\[\int dx\; T(x) f(x):=T(f)\]for a generalized function $T(x)$ and a function $f(x)\in {\cal A}$ with $x\in {\bf R}^4$. $\overline{z}$ designates the complex conjugate of a complex number $z$. The notation generally neglects to distinguish function sequences from component functions, for example, $f_n \in {\cal B}$ is an abbreviated designation for $f_n$ is a component function from a sequence $\underline{f}\in {\cal B}$.

The classical trajectories used in the descriptions of selected states are parametrized by a Newtonian time designated $\tau$ and the trajectories are designated ${\bf q}_k(\tau):=\hat{x}_k(\tau),\hat{y}_k(\tau),\hat{z}_k(\tau)$.

%= = = = = = = = = = = = = = = = = = = = = = =
% Minimum packet states
%= = = = = = = = = = = = = = = = = = = = = = =
\subsection{Minimum packets}\label{sec-minpak}

Functions that are strongly peaked near values that follow a classical trajectory determined by $1/r^2$ forces approximate in the short term solutions to a Schr\"odinger equation without a potential. The time in this Schr\"odinger equation is the temporal parameter of the classical trajectory.

From (\ref{B-defn}), spatial test functions generate labels of states of a relativistic UQFT. Test functions are also elements of ${\cal L}_2$. In ordinary (nonrelativistic) quantum mechanical descriptions, states are labeled by equivalence classes of ${\cal L}_2$ functions, time is an independent parameter, and Hermitian position and momentum observables satisfy the Heisenberg-Born-Jordan relation, $[P,X]=i \hbar$, realized in ${\cal L}_2$ by $X=x$ and $P=i\hbar \frac{d\;}{d x}$ with the convention (\ref{fourier}) for ${\cal L}_2$.

%= = = = = = the min packet function s = = = = = = = = = = = = = = = = =
${\cal L}_2$ minimum packet functions label quantum states that are most nearly described as classical states in the sense that the geometric means of the variances in position and momentum are minimal.\begin{equation} \label{minpak} \varphi({\bf x};m,{\bf q},\phi)=\frac{1}{(2\pi \sigma^2)^\frac{3}{4}}\; \exp\left(-\frac{({\bf x}-{\bf q})^2}{4\sigma^2}-i \frac{m \dot{\bf q}\!\cdot\!{\bf x}}{\hbar} +i\phi \right).\end{equation}This Schwartz function is described by the real mass $m$, classical trajectory ${\bf q}={\bf q}(\tau)$, and the real function $\phi=\phi(\tau)$. The time $\tau$ is a real argument of the functions ${\bf q}(\tau)$ and $\phi(\tau)$, and $\dot{\bf q}(\tau)$ is the first derivative with respect to $\tau$ of ${\bf q}(\tau)$. The Fourier transforms of these ${\cal L}_2$ minimum packets are\begin{equation} \label{minpak-p} \tilde{\varphi}({\bf p};m,{\bf q},\phi)=\left(\frac{2\sigma^2}{\pi}\right)^\frac{3}{4}\; \exp\left(-\sigma^2 ({\bf p}-\frac{m \dot{\bf q}}{\hbar})^2+i ({\bf p}-\frac{m \dot{\bf q}}{\hbar})\!\cdot\!{\bf q} +i\phi \right).\end{equation}For (\ref{minpak}) and with $\langle T \rangle := \int d{\bf x}\; \overline{\varphi({\bf x})}\, T\varphi({\bf x})$,\begin{equation}\renewcommand{\arraystretch}{1.25} \begin{array}{l} \langle 1\rangle= 1\\
\langle X\rangle= {\bf q}\\
\langle P\rangle= m \dot{\bf q}\\
\sigma_X^2=\langle (X-\langle X\rangle )^2 \rangle= \sigma^2\\
\sigma_P^2=\langle (P-\langle P\rangle)^2 \rangle= \hbar^2/(4\sigma^2)\\
\sigma_X \sigma_P =\hbar/2, \end{array}\end{equation}the minimum consistent with the Heisenberg uncertainty and the results are independent of ${\bf q}(\tau)$ and $\phi(\tau)$. The peak of the packet follows the trajectory ${\bf q}(\tau)$.

These minimum packet functions have properties of interest for the classical limits of states in UQFT.
\newline

{\em Lemma}: In a nonrelativistic, classical particle limit and over limited terms, the minimum packet functions (\ref{minpak}) approximate solutions to Schr\"{o}dinger's equation for a free particle with the Newtonian time parameter $\tau$.
\newline

The nonrelativistic, classical particle limit is described in the development below. For these minimum packet states (\ref{minpak}),\begin{equation}\label{d-varphi}\renewcommand{\arraystretch}{2.25} \begin{array}{rl} \dot{\varphi} &= ({\ds \frac{({\bf x}-{\bf q})\!\cdot\!\dot{\bf q}}{2\sigma^2}}-i {\ds \frac{m}{\hbar}} \ddot{\bf q}\!\cdot\! {\bf x}+i \dot{\phi})\,\varphi\\
\nabla \varphi &=(-{\ds \frac{({\bf x}-{\bf q})}{2\sigma^2}}-i {\ds \frac{m}{\hbar}} \dot{\bf q})\,\varphi\\
\nabla^2 \varphi &=(-{\ds \frac{1}{2\sigma^2}}+\left( {\ds \frac{({\bf x}-{\bf q})}{2\sigma^2}}+i {\ds \frac{m}{\hbar}} \dot{\bf q}\right)^2)\,\varphi.\end{array}\end{equation}$\nabla^2=\Delta$, the Laplacian in the three dimensions for ${\bf x}$ and $\dot{\varphi}$ designates the first derivative of $\varphi$ with respect to $\tau$.

Newton's equation of motion,\[m\ddot{\bf q} = F,\]and the derivatives of $\varphi$ from (\ref{d-varphi}) result in\begin{equation}\label{schro1} \renewcommand{\arraystretch}{2.25} \begin{array}{rl} i\hbar \dot{\varphi} &= (i\hbar{\ds \frac{({\bf x}-{\bf q})\!\cdot\!\dot{\bf q}}{2\sigma^2}}+m \ddot{\bf q}\!\cdot\! {\bf x}-\hbar \dot{\phi})\,\varphi\\
 &=\left({\ds \frac{\hbar^2}{2m}}\left({\ds \frac{({\bf x}-{\bf q})}{2\sigma^2}}+i {\ds \frac{m}{\hbar}} \dot{\bf q} \right)^2 -{\ds \frac{\hbar^2}{2m}}{\ds \frac{({\bf x}-{\bf q})^2}{4\sigma^4}} +{\ds \frac{m}{2}} \dot{\bf q}^2+ F\!\cdot\!{\bf x}-\hbar \dot{\phi}\right)\,\varphi\\
  &=\left({\ds \frac{\hbar^2}{2m}}\nabla^2 +{\ds \frac{\hbar^2}{2m}} \left({\ds \frac{1}{2\sigma^2}} -{\ds \frac{({\bf x}-{\bf q})^2}{4\sigma^4}}\right) +{\ds \frac{m}{2}} \dot{\bf q}^2+ F\!\cdot\!{\bf x}-\hbar \dot{\phi}\right)\,\varphi\end{array}\end{equation}from completion of the square. A definition\begin{equation}\label{serror}\epsilon({\bf x}-{\bf q}):= F\!\cdot\!({\bf x}-{\bf q})-{\ds \frac{\hbar^2}{2m}}{\ds \frac{({\bf x}-{\bf q})^2}{4\sigma^4}} \end{equation}results in\begin{equation}\label{schro2} i\hbar \dot{\varphi} =\left({\ds \frac{\hbar^2}{2m}}\nabla^2 +{\ds \frac{\hbar^2}{4m\sigma^2}} +{\ds \frac{m}{2}} \dot{\bf q}^2+ F\!\cdot\!{\bf q}+\epsilon({\bf x}-{\bf q})-\hbar \dot{\phi}\right)\,\varphi.\end{equation}Significantly, $\epsilon({\bf x}-{\bf q})$ is selected with\begin{equation}\label{exact}\epsilon({\bf 0})=0.\end{equation}

An energy is a constant of the classical motion.\[\hat{E}=\frac{m}{2} \dot{\bf q}^2+V(r)+\hat{E}_o\]in the notation from (\ref{conv-e}) of Appendix \ref{app-eqn-motion}. $\hat{E}$ designates the energy defined for the classical, nonrelativistic motion. For the $1/r^2$ force,\[F\!\cdot\!{\bf q}=V(r)\]and the definition of the energy results in\[ i\hbar \dot{\varphi} =\left({\ds \frac{\hbar^2}{2m}}\nabla^2 +{\ds \frac{\hbar^2}{4m\sigma^2}} +\hat{E}-\hat{E}_o+ \epsilon({\bf x}-{\bf q})-\hbar \dot{\phi}\right)\,\varphi\]from (\ref{schro2}). Setting\[\hbar \phi(\tau)=(mc^2+\hat{E}-\hat{E}_o+ {\ds \frac{\hbar^2}{4m\sigma^2}})\,\tau\]results in\begin{equation}\label{schro}-i\hbar \dot{\varphi} = (mc^2-{\ds \frac{\hbar^2}{2m}}\nabla^2-\epsilon({\bf x}-{\bf q})) \,\varphi.\end{equation}This is a Schr\"{o}dinger equation with a potential energy $\epsilon({\bf x}-{\bf q})$ and a time parameter $\tau$. The solution for $\phi$ is ambiguous up to an overall constant phase and the choice of $mc^2$ is natural for the nonrelativistic limit of the time evolution of UQFT states.

Before a study of solutions to this Schr\"{o}dinger equation, digress to demonstrate that for the $1/r^2$ force, $F\!\cdot\!{\bf q}=V(r)$. For a conservative, central force,\[F=-\nabla_{\!{\bf q}}V(r)\]with $r^2={\bf q}^2$ and the gradient is with respect to ${\bf q}$. From the chain rule for a central potential and with ${\bf q}=\hat{x},\hat{y},\hat{z}$,\[ \renewcommand{\arraystretch}{2.25} \begin{array}{rl} \nabla_{\!{\bf q}} V\!\cdot\!{\bf q} &= {\ds \frac{\partial V}{\partial \hat{x}}}\hat{x}+{\ds \frac{\partial V}{\partial \hat{y}}}\hat{y}+{\ds \frac{\partial V}{\partial \hat{z}}}\hat{z}\\
 &= {\ds \frac{dV}{dr}} ({\ds \frac{\partial r}{\partial \hat{x}}}\hat{x}+ {\ds \frac{\partial r}{\partial \hat{y}}}\hat{y}+{\ds \frac{\partial r}{\partial \hat{z}}}\hat{z})\\
 &= {\ds \frac{dV}{dr}} {\ds \frac{{\bf q}^2}{r}}\\
 &= {\ds \frac{dV}{dr}}\;r.\end{array}\]The solution to\[{\ds \frac{dV}{dr}}\;r=-V\]is $V(r)=a/r$ and includes the case of freely propagating particles, $V(r)=0$. This is the desired identity for the $1/r$ potential.

When $\epsilon({\bf x}-{\bf q})$ is negligible, the right hand side of this Schr\"{o}dinger equation (\ref{schro}) provides a nonrelativistic approximation to the UQFT Hamiltonian (\ref{hamil}). At the peak value of the functions (\ref{minpak}), ${\bf x}={\bf q}$ and $\epsilon({\bf x}-{\bf q})=0$. The states labeled by functions (\ref{minpak}) with a negligible $\epsilon({\bf x}-{\bf q})$ within the appreciable majority of the support of the functions are UQFT states with {\em nonrelativistic, classical limits}. Two conditions suffice to have a negligible $\epsilon({\bf x}-{\bf q})$. The conditions control the magnitudes of the two terms in (\ref{serror}). A negligible $\epsilon({\bf x}-{\bf q})$ applies in the approximation of solutions to the Schr\"{o}dinger equation for short term time evolution and within the peaked support of the functions. 

A lower bound on $\sigma$,\begin{equation}\label{nonrel}{\ds \frac{\hbar^2}{m\sigma^2}} \ll mc^2,\end{equation}partially controls the magnitude of $\epsilon({\bf x}-{\bf q})$ within the peaked support of the functions  (\ref{minpak}).\[0<{\ds \frac{\hbar^2}{2m}}{\ds \frac{({\bf x}-{\bf q})^2}{4\sigma^4}} < {\ds \frac{\hbar^2}{2m}}{\ds \frac{\lambda^2}{4\sigma^2}}\ll mc^2.\]$\|{\bf x}-{\bf q}\|<\lambda \sigma$ for a $\lambda>1$ within the most heavily weighted support of the functions (\ref{minpak}). The likelihood that $\|{\bf x}-{\bf q}\|<\lambda \sigma$ is more nearly certain for larger $\lambda$. This lower bound (\ref{nonrel}) provides a nonrelativistic limit from a limitation on the momentum support of the packets (\ref{minpak-p}). 

An upper limit on $\sigma$ is imposed to localize the support of the functions (\ref{minpak}). To accurately describe the evolution of the function as a particle, that is, to accurately describe the evolution of the function by the trajectory of its peak value, $\sigma$ is limited.\begin{equation}\label{limpak}|F\!\cdot\! ({\bf x}-{\bf q})| < \|F(r_a)\|\,\lambda \sigma \ll mc^2.\end{equation}The condition (\ref{limpak}) follows from the Cauchy inequality, that $\|F(r)\| <\|F(r_a)\|$ for a $1/r^2$ force with $r_a$ the distance $r$ at closest approach of the classical trajectories, and $\|{\bf x}-{\bf q}\|<\lambda \sigma$ for a $\lambda>1$ within the most heavily weighted support of the functions (\ref{minpak}). The $r_a$ are evaluated in Appendix \ref{app-eqn-motion} for the attractive $1/r^2$ force.

Satisfaction of the lower (\ref{nonrel}) and upper (\ref{limpak}) bounds provides that $\epsilon({\bf x}-{\bf q})$ is negligible with respect to $mc^2$ within the peaked support of the functions. Near the packet function peak, $|\epsilon({\bf x}-{\bf q})| \ll \hbar^2 \nabla^2/2m$  although as demonstrated in Section \ref{sec-schro} below, $\epsilon({\bf x}-{\bf q})$ is not necessarily negligible with regard to $\hbar^2 \nabla^2/2m$ toward the edges of the support under consideration.

The conditions (\ref{nonrel}) and (\ref{limpak}) are satisfied by packet spreads that are sufficiently large with respect to the Compton wavelength and sufficiently small that the potential energy difference across the packet at closest approach is much smaller than the rest mass energy (estimating with a contant potential across the packet).\[\frac{\hbar}{mc}\ll\; \sigma \ll\; \frac{mc^2}{\|F(r_a)\|}.\]In this case, the packet spreads that are compliant with the conditions (\ref{nonrel}) and (\ref{limpak}) describe the {\em nonrelativistic, classical limit} of this note. Satisfaction of the conditions imply that that the minimum packet functions centered on nonrelativistic, classical trajectories (\ref{minpak}) provide approximate solutions to Schr\"{o}dinger's equation in a neighborhood of every $\tau$ for selected parameters $r_a,\sigma$. The $r_a,\sigma$ can be selected to make the approximation arbitrarily precise when $m$ is arbitrarily large. $m\rightarrow \infty$ with $\sigma\rightarrow 0$ is a classical particle limit. For the example of gravity, $\|F(r_a)\|=Gm^2/r_a^2$, and setting\[\frac{\hbar^2}{m\sigma^2}=\frac{Gm^2}{r_a^2}\,\sigma \ll mc^2,\]closest approaches much greater than the Planck length,\[r_a \gg \left(\frac{\hbar G}{c^3}\right)^\frac{1}{2}=1.6\times 10^{-35}\mbox{ m}\]satisfy the condition (\ref{limpak}). In this case, the packet spreads that are compliant with the conditions (\ref{nonrel}) and (\ref{limpak}) are inversely proportional to mass.\[\sigma=\frac{1}{m}\left(\frac{\hbar^2r_a^2}{G}\right)^\frac{1}{3} = \frac{5.5\times 10^{-20}\,(r_a)^{\frac{2}{3}}}{m}\quad \mbox{kg-m}^{\frac{1}{3}}.\]

The result of the lemma follows from the selection of functions and is independent of the Wightman-functional. The lemma provides that in the short term for the nonrelativistic, classical limit, the evolution of the selected states follows the results of Newton's equations as demonstrated next.

%= = = = = = = = = = = = = = = = = = = = = = =
% Nonrelativistic limit equations of motion for classical particle states of UQFT
%= = = = = = = = = = = = = = = = = = = = = = =
\subsection{Equation of motion for nonrelativistic, classical limits of UQFT states} \label{sec-schro}

The evolution of states is determined by the local and Poincar\'{e} covariant Wightman-functional. For UQFT, over the short term for selected states in a nonrelativistic, classical limit, the peaked support of two particle states labeled by functions (\ref{minpak}) travel on trajectories described by Newtonian mechanics. This result is a specialized expansion of Ehrenfest's theorem to relativistic quantum physics for $1/r^2$ forces but the correspondence is quantum with classical state descriptions rather than a correspondence of Hermitian Hilbert space operators and classical dynamic quantities.

Using a separation of variables in Jacobi coordinates, two minimum packet states (\ref{minpak}) are used to label a state that describes two particles in a nonrelativistic, classical limit: one factor describes the motion of the center-of-mass; and one factor describes the relative motion. The commutation of derivatives applied to test functions provides that functions derived from (\ref{minpak}) that are elements of ${\cal B}$, described in (\ref{B-defn}), remain strongly peaked in the nonrelativistic limits. In nonrelativistic limits, the packet distortion due to the mapping from minimum packet to elements of ${\cal B}$ is dominated by low order derivatives. Symmetry of the UQFT functionals $W_n((x)_n)$ with interchange of arguments dissociates argument labels from particle labels. The particles may be indistinguishable.

From (\ref{B-defn}) and a separation of variables in Jacobi coordinates, (\ref{jacobi-factor}) in Appendix \ref{app-jacobi}, an element of ${\cal B}$ that is composed from minimum packet states (\ref{minpak-p}) is\begin{equation}\label{jacobi-b-factor}\tilde{f}_2(p_1,p_2) = (E_1+\omega_1)(E_2+\omega_2) \tilde{\varphi}({\ss \frac{1}{2}}({\bf p}_1-{\bf p}_2);\mu,{\bf q},\phi)\tilde{\varphi}({\bf p}_1+{\bf p}_2;m_T,{\bf q}_o,\phi_o).\end{equation}The appropriate mass for description of the relative motion is the reduced mass $\mu=m/2$ from (\ref{redcd-mass}) and the center of mass motion is described using $m_T=2m$ from  (\ref{total-mass}) in the case that $m:=m_1=m_2$. The description of the motion of the center-of-mass is a freely propagating trajectory ${\bf q}_o(\tau)$. That is,\[{\bf q}_o(\tau):={\bf a}+{\bf b}\,\tau\]for two constant 3-vectors, ${\bf a},{\bf b}$. 
\newline

\newcounter{theorems}
\setcounter{theorems}{1}
\renewcommand{\thetheorems}{\arabic{theorems}}
{\em Theorem} \thetheorems: In a nonrelativistic, classical limit of a UQFT and for the selected states (\ref{jacobi-b-factor}), short term state evolution follows trajectories given by Newton's equation with a $1/r^2$ force.
\newline

This $\tilde{f}_2(p_1,p_2)\in {\cal B}$ from (\ref{jacobi-b-factor}) describes a two argument state with a nonrelativistic, classical limit as two particles coupled by an attractive $1/r^2$ force. The nonrelativistic limits are developed for these elements from ${\cal B}$ with point support in time and $t_j=t$ for each argument $x_j$. The energies that result in evaluation of $\phi+\phi_o$ derive from (\ref{energy-jacobi}) of Appendix \ref{app-eqn-motion},\begin{equation}\label{ehat}\renewcommand{\arraystretch}{2.25} \begin{array}{rl} \hat{E} &= {\ds \frac{1}{2}} m_T\|\dot{\bf q}_o\|^2 + {\ds \frac{1}{2}} \mu \|\dot{\bf q}\|^2 +V({\bf q})+\hat{E}_o\\
 &= {\ds \frac{1}{2}} m\|\dot{\bf q}_1\|^2 + {\ds \frac{1}{2}} m \|\dot{\bf q}_2\|^2 +V({\bf q}_1-{\bf q}_2)+\hat{E}_o.\end{array}\end{equation}

As a consequence of the weighted support of the functions (\ref{minpak-p}),\[\renewcommand{\arraystretch}{2.25} \begin{array}{rl}\omega_j &= \sqrt{\left({\ds \frac{mc}{\hbar}}\right)^2+{\bf p}_2^2\;}\\
 &\approx {\ds \frac{1}{\hbar c}}\left(mc^2 + {\ds \frac{\hbar^2}{2m}} {\bf p}_j^2\right)\end{array}\]from\[{\ds \frac{\hbar^2}{2m}} {\bf p}_j^2\ll mc^2\]for $j=1,2$. These upper bounds are a consequence of condition (\ref{nonrel}). The support of the functions (\ref{minpak-p}) are peaked and heavily weighted within\[\frac{\hbar^2}{2m_T} ({\ss \frac{1}{2}}({\bf p}_1-{\bf p}_2)-\frac{m_T \dot{\bf q}_o}{\hbar})^2< \frac{\hbar^2}{2m_T} \frac{\lambda^2}{\sigma^2}\]and\[\frac{\hbar^2}{2\mu} ({\bf p}_1+{\bf p}_2-\frac{\mu \dot{\bf q}}{\hbar})^2< \frac{\hbar^2}{2\mu} \frac{\lambda^2}{\sigma^2}.\]$\lambda> 1$. Then, the condition (\ref{nonrel}),\[\frac{\hbar^2}{2m} \frac{\lambda^2}{4\sigma^2}\ll mc^2,\]the nonrelativistic motion of the classical trajectories\[\dot{\bf q}_o\ll c\qquad \mbox{and}\qquad \dot{\bf q}\ll c\]and these weighted contributions of the functions (\ref{minpak-p}) provide that\[{\ds \frac{\hbar^2}{2m}}{\bf p}_1^2 +{\ds \frac{\hbar^2}{2m}}{\bf p}_2^2 \ll mc^2\]within the peaked support of the state labeled by (\ref{jacobi-b-factor}).

Finally, the separation of variables in Jacobi coordinates, the form of the UQFT Hamiltonian (\ref{hamil}), the nonrelativistic approximation, and the lemma result in\begin{equation}\label{local-schro} \renewcommand{\arraystretch}{2.25} \begin{array}{rl}i\hbar {\ds \frac{\partial f_2}{\partial t}} &=\left(\sqrt{m^2c^4 -\hbar^2c^2 \nabla_1^2}+ \sqrt{m^2c^4 -\hbar^2c^2 \nabla_2^2}\right) f_2\\
 &\approx \left(2mc^2-{\ds \frac{\hbar^2}{2m}}\nabla_1^2-{\ds \frac{\hbar^2}{2m}}\nabla_2^2\right) f_2\\
 &\approx -i\hbar \,\dot{f}_2\end{array}\end{equation}with $f_2(x_1,x_2)$ the inverse Fourier transform of (\ref{jacobi-b-factor}).

With the concentrated support of $f_2$, the approximation $\partial/\partial t\approx -\partial/\partial \tau$ in (\ref{local-schro}) connects the quantum dynamics to the nonrelativistic, classical particle trajectories. $\partial/\partial t$ generates displacements of states in time and for the selected states, $\partial/\partial\tau$ generates displacements in the temporal parameter describing classical trajectories. The support of the selected states (\ref{jacobi-b-factor}) is concentrated near the classical trajectories. The sign difference is due to the convention that $U(-\tau) f(0)=f(\tau)$ from $U(t)f(t')=f(t'-t)$. With $\underline{f}=(0,0,f_2,0\ldots)$ and $f_2$ from (\ref{jacobi-b-factor}), the result is that\begin{equation}\label{approx1}U(\delta t) |\underline{f}\rangle \approx (1- \delta t\; \hbar \frac{\partial\;}{\partial t}) |\underline{f}\rangle \approx (1+ \delta t\; \hbar \frac{\partial\;}{\partial \tau}) |\underline{f}\rangle.\end{equation}The two argument functions (\ref{jacobi-b-factor}) label states that exhibit interaction. When $\hat{E} <\hat{E}_o$ and the force is attractive, the classical trajectories are bound and the range separation ${\bf q}$ is finite for all $\tau$.

The approximation (\ref{approx1}) applies for the selected states labeled by functions (\ref{jacobi-b-factor}). For this same subset of states, the support of the states follows classical trajectories for $1/r^2$ forces in appropriate nonrelativistic, classical limits in the short term. This association substitutes for the conjecture of a canonical quantization, that there is a correspondence of operators with the classical equations of motion.

The theorem follows for the $1/r$ potential and from: the form of the UQFT Hamiltonian (\ref{hamil}); the form of the functions that label the selected states (\ref{jacobi-b-factor}); and that interaction is exhibited (momenta are not constrained to be equal in pairs). The result is otherwise independent of the form of the Wightman-functional. In the long term, exhibition of a $1/r$ potential in nonrelativistic limits of the scattering cross sections does constrain the form of the Wightman-functional. The scattering cross sections for a potential are calculated for distinguishable particles and different limits of the functions (\ref{minpak}), the plane wave and long term limit. Plane wave limits are included in the parametric range of (\ref{minpak}) as the case with divergent spatial packet spread. A Wightman-functional satisfying the constraint to exhibit a $1/r$ potential in the scattering cross section is provided in [\ref{feymns}]. A first Born approximation, mollified $1/r$ potential scattering cross section agrees with the classical Rutherford and the exact $1/r$ potential calculations [\ref{rgnewton}]. Corrections to the equivalent $1/r$ potential at very small and very great distances are sufficient to achieve a continuous linear Wightman-functional. 

The long term time evolution of states is determined by the Wightman-functional. The local and Poincar\'{e} covariant UQFT Wightman-functional provides that state evolution satisfies the approximation (\ref{approx1}) in the short term for the selected states and the nonrelativistic, classical limits (\ref{nonrel}) with (\ref{limpak}). It has not been demonstrated that UQFT reduces to ordinary quantum mechanics in a nonrelativistic limit although the evaluation of the connected four-point functional in the scattering calculation [\ref{feymns}] provides that the effective interaction is approximately a $1/r$ potential in a nonrelativistic limit. For the example of the hydrogen atom, an electron in a bound state does not satisfy the lower bound on $\sigma$ (\ref{nonrel}) for the classical limit to apply.

%= = = = = = = = = = = = = = = = = = = = = = =
% Concluding remarks
%= = = = = = = = = = = = = = = = = = = = = = =
\section{Concluding remarks}

Wightman-functionals can be selected to exhibit a nonrelativistic, classical limit with two particle states interacting with a $1/r^2$ mutual force and with nonrelativistic limit elastic scattering cross sections that coincide with the results from ordinary quantum mechanics. The explicit, relativistic, local UQFT are well approximated by Newtonian gravity in the short term and in a nonrelativistic, classical limit. 

UQFT conforms to both the principles of special relativity and quantum mechanics. The resulting Wightman-functional is a description of local quantum physics that includes cases of evident physical interest. UQFT escapes the inconsistencies of established relativistic QFT [\ref{pct},\ref{bogo},\ref{jost},\ref{wightman-hilbert}] by considering only the Hilbert space operators that are necessarily present. These operators include the orthogonal projections onto subspaces of states and the generators of Poincar\'{e} symmetries. Conjecture concerning correspondence of classical dynamic quantities with self-adjoint Hilbert space operators is excluded. In rigged (equipped) Hilbert spaces (Gelfand triples), observable quantities include the summation variables in representations of generalized functions as weighted summations, a concept that includes both energy-momentum and spacetime coordinates [\ref{johnson}]. The UQFT generalization to conventional quantum mechanics is based on making fewer assumptions and, if observables were necessarily Hermitian Hilbert space operators, then we would be inevitably led to this conclusion by the physical requirements. However, this is not the case for the UQFT constructions and [\ref{wigner}] stands as a counterexample to the conjectured correspondence. Canonical quantization has made correspondence of classical dynamic quantities with self-adjoint Hilbert space operators a foundational principle of QFT despite an obscure physical foundation. Unification of relativity with quantum mechanics without this disputable assertion departs from the challenges posed by inapplicability of the Stone-von Neumann theorem, the implications of the Haag and Hall-Wightman-Greenberg theorems, and the artifacts of conjectured representations for interaction ``at a point''.

This revisit to the foundations of quantum mechanics to unify relativity with quantum mechanics when interaction is exhibited follows the thought expressed by David Hilbert [\ref{wightman-hilbert}]:\begin{quote}I believe that specialization plays an even more important role than generalization when one deals with mathematical problems. Perhaps in most cases in which we seek in vain the answer to a question, the cause of failure lies in the fact that we have worked out simpler and easier problems either not at all or incompletely. What is important is to locate these easier problems and to work out their solutions with tools that are as complete as possible and with concepts capable of generalization.\end{quote} 

%= = = = = = = = = = = = = = = = = = = = = = =
%= = = = = = = Appendices  = = = = = = = = = =
%= = = = = = = = = = = = = = = = = = = = = = =
\section*{Appendices}
\renewcommand{\thesubsection}{\Alph{subsection}}
\setcounter{subsection}{0}

%= = = = = = = = = = = = = = = = = = = = = = =
% Jacobi coordinates
%= = = = = = = = = = = = = = = = = = = = = = =
\subsection{Jacobi coordinates}\label{app-jacobi}

The equations of motion for the relative and center-of-mass motions of two particles separate in Jacobi coordinates. When ${\bf x}_1,{\bf x}_2$ are the spatial coordinates of two particles, then\begin{equation}\label{jacobi} {\bf q}:={\bf x}_1-{\bf x}_2 \qquad \qquad \mbox{and}\qquad \qquad {\bf q}_o:= {\ds \frac{m_1{\bf x}_1+m_2{\bf x}_2}{m_1+m_2}}\end{equation}are the Jacobi coordinates and ${\bf q}_o=\frac{1}{2}({\bf x}_1+{\bf x}_2)$ when $m_1=m_2$. An ${\cal L}_1 \cap {\cal L}_2$ function that separates in Jacobi coordinates,\begin{equation}\label{jacobi-factor-st}f_2({\bf x}_1,{\bf x}_2) = \varphi({\bf x}_1-{\bf x}_2)\psi(\frac{m_1{\bf x}_1+m_2{\bf x}_2}{m_1+m_2}),\end{equation}has a Fourier transform\begin{equation}\label{jacobi-factor}\tilde{f}_2({\bf p}_1,{\bf p}_2) = \tilde{\varphi}(\frac{m_2 {\bf p}_1-m_1 {\bf p}_2}{m_1+m_2})\tilde{\psi}({\bf p}_1+{\bf p}_2).\end{equation}The Jacobi coordinates invert as\begin{equation}\label{jacobi-inv} {\bf x}_1={\bf q}_o+ \frac{m_2{\bf q}}{m_1+m_2} \qquad \qquad \mbox{and}\qquad \qquad {\bf x}_2={\bf q}_o-\frac{m_1 {\bf q}}{m_1+m_2}.\end{equation}

The Jacobi coordinates provide that if $\varphi$ and $\psi$ satisfy equations of motion\begin{equation}\label{eqn-decouple}\renewcommand{\arraystretch}{1.25} \begin{array}{rl} (-{\ds \frac{\hbar^2}{2\mu}}\nabla_{{\bf q}}^2 +V({\bf q}))\varphi({\bf q},t) &=i\hbar {\ds \frac{\partial\varphi({\bf q},t)}{\partial t}}\\
 -{\ds \frac{\hbar^2}{2m_T}}\nabla_{{\bf q}_o}^2 \psi({\bf q}_o,t) &=i\hbar {\ds \frac{\partial \psi({\bf q}_o,t)}{\partial t}},\end{array}\end{equation}then the chain rule of diffentiation provides a solution to\begin{equation}\label{eqn-of-motn} (-\frac{\hbar^2}{2m_1}\nabla_{{\bf x}_1}^2 -\frac{\hbar^2}{2m_2}\nabla_{{\bf x}_2}^2 +V({\bf x}_1-{\bf x}_2))f({\bf x}_1,{\bf x}_2,t) =i\hbar \frac{\partial f({\bf x}_1,{\bf x}_2,t)}{\partial t}\end{equation}by separation of variables,\begin{equation}\label{sep-varbls} f({\bf x}_1,{\bf x}_2,t) =\varphi({\bf x}_1-{\bf x}_2,t) \psi(\frac{m_1{\bf x}_1+m_2{\bf x}_2}{m_1+m_2},t).\end{equation}In (\ref{eqn-decouple}),\begin{equation}\label{redcd-mass}\mu:=\frac{m_1m_2}{m_1+m_2}\end{equation}is the reduced mass and\begin{equation}\label{total-mass}m_T:=m_1+m_2\end{equation}is the total mass. The derivatives are partial derivatives with the components of ${\bf x}$ and $t$ considered as independent variables.\[\nabla_{\bf x}^2=\Delta_{\bf x}:= \frac{\partial^2\;}{\partial x^2}+\frac{\partial^2\;}{\partial y^2}+\frac{\partial^2\;}{\partial z^2}\]is the Laplacian in three dimensions.

%= = = = = = = = = = = = = = = = = = = = = = =
% Classical eqn of motion, F=ma, in Jacobi coordinates
%= = = = = = = = = = = = = = = = = = = = = = =
\subsection{Newton's equations of motion}\label{app-eqn-motion}

The particle trajectories used to define the functions (\ref{minpak}) are described by Newton's principles that result in the equation of motion, $F=m{\ddot{\bf q}}$. The trajectories are parametrized by $\tau$, ${\bf q}={\bf q}(\tau)$.

In classical mechanics, the states of two particles are vectors ${\bf q}_k:=\hat{x}_k,\hat{y}_k,\hat{z}_k$ in a three dimensional Euclidean configuration space and the locations are functions parameterized by time. With the Euclidean distance\[r:=\| {\bf q}_1-{\bf q}_2\|\]and the unit vector pointing from particle 2 to particle 1\[{\bf u}_r:= \frac{{\bf q}_1-{\bf q}_2}{r},\]the equations of motion for the two particles are\begin{equation}\label{motion1} F = m_1 \ddot{{\bf q}}_1\qquad \qquad \mbox{and}\qquad \qquad -F = m_2 \ddot{{\bf q}}_2.\end{equation}The signs to implement an attractive, central force are $F=-\|F\|\,{\bf u}_r$, each particle is drawn towards the other. \begin{equation}\label{f-grav} F=-{\ds \frac{Gm_1m_2}{r^2}} {\bf u}_r\end{equation}is Newtonian gravity.

The Jacobi coordinates (\ref{jacobi}) for the classical trajectory,\[ {\bf q}:={\bf q}_1-{\bf q}_2 \qquad \qquad \mbox{and}\qquad \qquad {\bf q}_o:= {\ds \frac{m_1{\bf q}_1+m_2{\bf q}_2}{m_1+m_2}},\]describe the relative motion and the motion of the center-of-mass, respectively. The equations of motion (\ref{motion1}) become\begin{equation}\label{motion2} \ddot{{\bf q}}_o=0\qquad \qquad \mbox{and}\qquad \qquad \mu \ddot{{\bf q}}= F\end{equation}from\[\renewcommand{\arraystretch}{1.75} \begin{array}{rl} \ddot{\bf q}_1-\ddot{\bf q}_2 &= F \;\left({\ds \frac{1}{m_1}} +{\ds \frac{1}{m_2}}\right) = {\ds \frac{1}{\mu}}\, F\\
m_1\ddot{\bf q}_1+m_2\ddot{\bf q}_2 &=0\end{array}.\]with the reduced mass $\mu$ from (\ref{redcd-mass}) and total mass $m_T$ from (\ref{total-mass}). The relative range may be bounded depending upon the energies of the incoming particles.

For a conservative force,\[F=-\nabla V({\bf q}),\] and $F=\mu \ddot{\bf q}$ results in\[(\mu \ddot {\bf q}-F) \cdot \dot{\bf q}=\mu \ddot {\bf q}\cdot \dot {\bf q}+\nabla V({\bf q}) \cdot \dot{\bf q}=\frac{d\;}{d\tau} \left( \frac{\mu}{2} \|\dot{\bf q}\|^2 +V({\bf q}) \right) =0\]using the chain rule for derivatives. With a conservative force, the total energy is a constant of motion,\begin{equation}\label{conv-e}\hat{E}:=\frac{1}{2} \mu \|\dot{\bf q}\|^2 +V({\bf q})+\hat{E}_o.\end{equation}$\hat{E}_o$ is selected to set $\hat{E}\geq 0$ but the attractive $1/r^2$ force corresponds to an infinitely deep potential. In this case, no finite constant necessarily results in a nonnegative energy. A regularization of $V({\bf q})$ is required to ensure that $\hat{E}\geq 0$ with a finite $\hat{E}_o$.

In the multiple body case, the energies add. In the two body case without an external potential,\begin{equation}\label{energy}\hat{E}:=\frac{1}{2} m_1\|\dot{\bf q}_1\|^2 +\frac{1}{2} m_2\|\dot{\bf q}_2\|^2 +V({\bf q}_1-{\bf q}_2)+\hat{E}_o.\end{equation}In Jacobi coordinates,\begin{equation}\label{energy-jacobi}\renewcommand{\arraystretch}{2.25} \begin{array}{rl} \hat{E}&= {\ds \frac{m_1(m_1+m_2)\|\dot{\bf q}_1\|^2 + m_2(m_1+m_2)\|\dot{\bf q}_2\|^2}{2(m_1+m_2)}} +V({\bf q}_1-{\bf q}_2)+\hat{E}_o\\
 &= {\ds \frac{\|m_1\dot{\bf q}_1+m_2\dot{\bf q}_2\|^2 + m_1m_2\|\dot{\bf q}_1-\dot{\bf q}_2\|^2}{2(m_1+m_2)}} +V({\bf q}_1-{\bf q}_2)+\hat{E}_o\\
 &= {\ds \frac{1}{2}} m_T\|\dot{\bf q}_o\|^2 + {\ds \frac{1}{2}} \mu \|\dot{\bf q}\|^2 +V({\bf q})+\hat{E}_o.\end{array}\end{equation}

The central $1/r^2$ potential is conservative.\[\renewcommand{\arraystretch}{1.75} \begin{array}{rl} F&= -{\ds \frac{Gm_1m_2}{r^2}\,\frac{\bf q}{r}} = Gm_1m_2 \nabla {\ds \frac{1}{r}}\\
  &= - \nabla V\end{array}\]with\begin{equation}\label{potential}V = -{\ds \frac{Gm_1m_2}{r}}.\end{equation}

Solutions for the relative motion ${\bf q}$ lie in a plane as a consequence of a force that lies in the plane defined by the particles' initial relative position and initial relative velocity. As a result, the motions never leave that plane. In a coordinate system with\begin{equation}\label{coords} {\bf q}=(\hat{x},\hat{y},0),\end{equation}polar coordinates are\begin{equation}\label{polar-coords} \hat{x}:=r\, \cos \theta \qquad \mbox{and}\qquad \hat{y}=r \sin \theta.\end{equation}The equation of motion for the relative motion (\ref{motion2}) results in\[\renewcommand{\arraystretch}{2.25} \begin{array}{rl} \ddot{\hat{x}}&= \ddot{r} \cos \theta -r \dot{\theta}^2 \cos \theta -2\dot{r}\dot{\theta} \sin \theta -r\ddot{\theta} \sin \theta = -{\ds \frac{Gm_T}{r^2}}\cos \theta\\
\ddot{\hat{y}}&= \ddot{r} \sin \theta -r \dot{\theta}^2 \sin \theta +2\dot{r}\dot{\theta} \cos \theta +r\ddot{\theta} \cos \theta = -{\ds \frac{Gm_T}{r^2}}\sin \theta\end{array}\]with $m_T$ the total mass $m_1+m_2$ from (\ref{total-mass}). This results in\[\ddot{r}-r \dot{\theta}^2 = -\frac{Gm_T}{r^2}\]and\[2\dot{r}\dot{\theta}+r \ddot{\theta} = 0.\]This second relation provides that\[\frac{d\;}{d\tau}(r^2 \dot{\theta})=0\]or that $L:=r^2 \dot{\theta}$ is a constant of the motion, an angular momentum divided by the reduced mass $\mu$. Using this relation, the second equation of motion results in\begin{equation}\label{motion3} \ddot{r}-\frac{L^2}{r^3} = -\frac{Gm_T}{r^2}\end{equation}and\[\frac{1}{2} \|\dot{\bf q}\|^2= \frac{1}{2} \dot{r}^2 +\frac{1}{2} \frac{L^2}{r^2}.\]

Solutions to the equations of motion (\ref{motion2}) describe trajectories that are conic sections and the solutions can be divided into the scattering solutions and the bound state solutions by the value of $\hat{E}-\hat{E}_o$ from (\ref{energy}). The scattering solutions have $\hat{E}>\hat{E}_o$ and bound states result when $\hat{E}<\hat{E}_o$.\[r=r(\theta)\qquad \mbox\qquad \tau=\tau(\theta)\]with, in the example of the $1/r^2$ force,\[\hat{x}=r \cos \theta \qquad \mbox{and}\qquad \hat{y}=r \sin\theta\]and\begin{equation}\label{clost-approach}r(\theta) = \left\{ \renewcommand{\arraystretch}{2.25} \begin{array}{lll} {\ds \frac{b^2}{a-\sqrt{a^2+b^2}\,\cos\theta}}&\geq \sqrt{a^2+b^2}-a &\qquad \mbox{scattered solution, }\hat{E}>\hat{E}_o\\
 {\ds \frac{b^2}{a-\sqrt{a^2-b^2}\,\cos\theta}}&\geq a-\sqrt{a^2-b^2} &\qquad \mbox{bound state, }\hat{E}<\hat{E}_o\\
 {\ds \frac{2a}{1-\cos\theta}}&\geq a &\qquad \mbox{transition solution, }\hat{E}=\hat{E}_o\end{array}\right.\end{equation}The Newtonian time $\tau$ for each point in the trajectory is provided by\[\tau(\theta) = \frac{1}{L} \int_{\theta_0}^{\theta} ds\; r(s)^2\]$\tau(\theta_0):=0$ and $L=r^2 \dot{\theta}$ is a constant of the motion.

The scattered solutions are hyperbolas with both foci on the $x$-axis, one focus at the origin, closest approach to origin of $\sqrt{a^2+b^2}-a$ for the right-hand curve segment used and the slopes of the asymptotes are $\pm b/a$. For the scattered solutions $-\cos^{-1}(\frac{a}{\sqrt{a^2+b^2}}) \leq \theta < 2\pi-\cos^{-1}(\frac{a}{\sqrt{a^2+b^2}})$. The bound state solutions are ellipses with both foci on the $x$-axis, major axis $2a$ and minor axis $2b$, $0 \leq \theta \leq 2\pi$ and $b^2<a^2$. The transition solutions are parabolas with the focus at the origin, both the vertex and focus on the $x$-axis, and $0 \leq \theta \leq 2\pi$. For the scattered and bound state trajectories,\[a=\frac{Gm_1m_2}{2\,|\hat{E}-\hat{E}_o|}\qquad \mbox \qquad b^2=\frac{\mu L^2}{2\,|\hat{E}-\hat{E}_o|}\]and for the transition trajectories,\[a=\frac{L}{2(m_1+m_2)G}.\]

%= = = = = = = = = = = = = = = = = = = = = = =
% a single, neutral, Lorentz scalar field
%= = = = = = = = = = = = = = = = = = = = = = =
\subsection{UQFT for a single, neutral, Lorentz scalar field}\label{app-uqft}

Unconstrained QFT [\ref{gej05}] are a generalization of the Wightman-functional development of QFT [\ref{wight},\ref{borchers}]. The properties of a quantum field are described by a Wightman-functional $\underline{W}$ that is a sequence of generalized functions dual to functions in the linear topological space ${\cal A}$ consisting of sequences of functions $\underline{f}$ with Fourier transforms that are Schwartz tempered test functions of the momenta ${\bf p}_j\in {\bf R}^3$ when energies are evaluated on mass shells, $E_j=\pm \omega_j$. The sesquilinear function on ${\cal A}\times {\cal A}$,\begin{equation}\label{sesquis}\underline{W}(\underline{f}^* \,{\bf x}\, \underline{g}) :={\ds \sum_{n,m} \int} d(p)_{n+m}\;\tilde{W}_{n+m}((p)_{n+m}) \overline{\widetilde{f}_n}((-p)_{n,1})\, \tilde{g}_m((p)_{n+1,n+m})\end{equation}provides the scalar product for elements in the Hilbert space. $\underline{W}(\underline{f}^* \,{\bf x}\, \underline{f})\geq 0$ for $\underline{f}\in {\cal B}$. The Hilbert space representation of states is the result of a bijective map of equivalence classes of elements $\underline{f}\in {\cal B}$ from (\ref{B-defn}) for the semi-norm\begin{equation}\label{norm} \| \underline{f} \|_{\cal B}:= \sqrt{\underline{W}(\underline{f}^* \, {\bf x}\, \underline{f})}\end{equation}to a dense set of elements in the Hilbert space. This map,\begin{equation}\label{isometry}\langle \underline{f}|\underline{g}\rangle= \underline{W}(\underline{f}^* \,{\bf x}\, \underline{g})\end{equation}is an isometry.

For $f_n,g_m \in {\cal B}$, the connected functions for a single, neutral, Lorentz scalar field are\[{^CW}_2(f_1^* g_1)=W_2(f_1^* g_1)=\Delta(f_1^* g_1)\]for the positive frequency Pauli-Jordan function $\Delta(x_1-x_2)$ and when $n+m\geq 3$,\begin{equation}\label{connctd-eval} \renewcommand{\arraystretch}{1.25} \begin{array}{l} {^CW}_{n+m}(f_n^*\, g_m) := \overline{\varsigma_n}\, \varsigma_m \, c_{n+m}{\ds \int} (dp)_{n+m}{\ds \int} du\; {\ds \prod_{k=1}^{n+m} \exp(is_k p_k u)\, \delta^+_k \left(\frac{\ds \partial\;}{\ds \partial\rho_k} \right)}\times\\
 \quad \exp\left({\ds \sum_{i<j}^n} \rho_i \rho_j \,\overline{U_n(p_i\!-\!p_j)}+\!{\ds \sum_{n<i<j}^{n+m}} \rho_i \rho_j \,U_m(p_i\!-\!p_j) +\!{\ds \sum_{i=1}^n \sum_{j=n+1}^{n+m}} \rho_i \rho_j \beta_{i+j-n}\Upsilon(p_i\!+\!p_j) \right)\times\\
 \quad  \hat{\bf S}[ \overline{\tilde{f}_n}((p)_n)]\,\hat{\bf S}[\tilde{g}_m((p)_{n+1,n+m})]\end{array} \end{equation}evaluated at $(\rho)_{n+m}=0$. $s_k=-1$ for $k \leq n$ and $s_k=1$ otherwise. $\hat{\bf S}[]$ indicates summation over all $n!$ distinct permutations for the order of the $n$ arguments normalized by the number of permutations, $n!$. In particular,\[\hat{\bf S}[\tilde{f}_2(p_1,p_2)]=\frac{1}{2!}(\tilde{f}_2(p_1,p_2)+\tilde{f}_2(p_2,p_1)).\]These connected functions ${^CW}_n$ are identified as the connected contributions [\ref{gej05}] of the $n$-point Wightman functionals $W_n$. The $n$-point Wightman functionals are generalized functions composed of finite sums of products of connected functions without arguments in common and described by:\begin{list}{}{\itemsep -0.06in} \item[--] the Pauli-Jordan, free field two-point function $\Delta(x_1\!-\!x_2)$ that determines the single elementary particle of mass $m$ \item[--] coefficients $c_n$ that are the moments of a nonnegative measure\[c_n:=\int \sigma(d\lambda)\; \lambda^n\] \item[--] complex constants $\varsigma_n$ and $\varsigma_2=1$ without loss of generality \item[--] Lorentz invariant functions $U_n(p),\Upsilon(p)$ that are multipliers of tempered functions and $\Upsilon(p)$ is the Laplace transform of a nonnegative measure\[\Upsilon(p) ={\ds \int} d\mu_u(s)\; e^{-sp}\] \item[--] coefficients $\beta_j$ that are Laplace transforms of a nonnegative measure\[\beta_j:={\ds \int} \mu_\beta(dv)\; e^{-jv}.\]\end{list}Variations of these constructions include convex sums of the connected functions (\ref{connctd-eval}) and additional organizations for Lorentz invariant functions into nonnegative forms [\ref{feymns}].

%= = = = = = = 
The four-point functions are the sum of a free field four-point function and the four-point connected function (\ref{scalar-amp}). For $W_4(f_2^* g_2)$ and $f_2, g_2\in {\cal B}$, the free field contribution to the four-point function results in\begin{equation}\label{link4} \tilde{W}_4((p)_4) = {^C\tilde{W}_4((p)_4)} +\tilde{\Delta}_{12}\tilde{\Delta}_{34}+\tilde{\Delta}_{14}\tilde{\Delta}_{23} +\tilde{\Delta}_{13}\tilde{\Delta}_{24}\end{equation}with the Fourier transform of a Pauli-Jordan function\[ \tilde{\Delta}_{ij}:=\tilde{\Delta}(p_i,p_j) := \delta(p_i+p_j)\delta_j^+\]with $\delta_j^+:= \theta(E_j)\delta(p_j^2-m^2)$. $\tilde{\Delta}_{12}\tilde{\Delta}_{34}$ does not contribute when $f_2,g_2 \in {\cal B}$. For $f_2,g_2 \in {\cal B}$, evaluation of (\ref{connctd-eval}) results in the four-point connected function\begin{equation}\label{scalar-amp} \renewcommand{\arraystretch}{1.75} \begin{array}{l} {^CW}_4(f_2^*\, g_2) := c_4{\ds \int} (dp)_4{\ds \int} du\; {\ds \prod_{k=1}^{4} \left(\frac{\delta^+_k\, e^{is_k p_k u}}{2\omega_k}\right)}\; \overline{\tilde{f}_2}(p_1,p_2)\,\tilde{g}_2(p_3,p_4)\times\\
\hat{{\bf S}}_{1,2}[\hat{{\bf S}}_{3,4}[\overline{U_2(p_1\!-\!p_2)}\,U_2(p_3\!-\!p_4)\!+\!\beta_2\Upsilon(p_1\!+\!p_3) \beta_4\Upsilon(p_2\!+\!p_4)\!+\!\beta_3\Upsilon(p_1\!+\!p_4) \beta_3\Upsilon(p_2\!+\!p_3)]]\\
 \qquad = c_4{\ds \int} (dp)_4{\ds \int} du\; {\ds \prod_{k=1}^{4} \left(\frac{\delta^+_k\, e^{is_k p_k u}}{2\omega_k}\right)}\; \overline{\tilde{f}_2}(p_1,p_2)\,\tilde{g}_2(p_3,p_4)\, \left[
\overline{U_e(p_1\!-\!p_2)}\,U_e(p_3\!-\!p_4)\right.\\
 \qquad \qquad +\frac{1}{2}(\beta_2\beta_4+\beta_3^2)\Upsilon(p_1\!+\!p_3) \, \Upsilon(p_2\!+\!p_4)\left. + \frac{1}{2}(\beta_2\beta_4+\beta_3^2)\Upsilon(p_1\!+\!p_4) \, \Upsilon(p_2\!+\!p_3) \right]\end{array} \end{equation}and\[2\, U_e(p):=U_2(p)+U_2(-p)\]is the even contribution of $U_2(p)$. $\hat{\bf S}_{j,k}[]$ is the normalized symmetrization of arguments $p_j$ through $p_k$ with respect to argument order. Evaluation of (\ref{link4}) results in the free field contribution to the four-point function\begin{equation}\label{scalar-free} \renewcommand{\arraystretch}{1.25} \begin{array}{rl} W_4(f_2^*\, g_2)-{^CW}_4(f_2^*\, g_2) &= {\ds \int} (dp)_4{\ds \int} {\ds \prod_{k=1}^{4} \frac{\delta^+_k}{\sqrt{2\omega_k}}}\; \overline{\tilde{f}_2}(p_1,p_2)\,\tilde{g}_2(p_3,p_4)\times\\
 &\qquad\qquad\qquad \left(\delta({\bf p}_1\!-\!{\bf p}_3)\delta({\bf p}_2\!-\!{\bf p}_4)+\delta({\bf p}_1\!-\!{\bf p}_4)\delta({\bf p}_2\!-\!{\bf p}_3)\right) \end{array} \end{equation}
%= = = = = = = = = = = = = = = = = = = = = = =
%= = = = = = = = References= = = = = = = = = =
%= = = = = = = = = = = = = = = = = = = = = = =
\section*{References}
\begin{enumerate}
\item \label{gej05} G.E.~Johnson, ``Algebras without Involution and Quantum Field Theories'', March 2012, arXiv:math-ph/1203.2705. And, G.E.~Johnson, ``Introduction to quantum field theories exhibing interaction'', August 2014, to appear.
\item \label{wight} A.S.~Wightman, ``Quantum Field Theory in Terms of Vacuum Expectation Values'', {\em Phys.~Rev.}, vol.~101, 1956, p.~860.
\item \label{borchers} H.J.~Borchers, ``On the structure of the algebra of field operators'', {\em Nuovo Cimento}, Vol.~24, 1962, p.~214.
\item \label{johnson} G.E.~Johnson, ``Measurement and self-adjoint operators'', May 2014, arXiv:quant-ph/\-1405.\-7224.
\item \label{feymns} G.E.~Johnson, ``Fields and Quantum Mechanics'', Dec.~2013, arXiv:math-ph/\-1312.\-2608.
\item \label{gel2} I.M.~Gel'fand, and G.E.~Shilov, {\em Generalized Functions}, Vol.~2, trans.~M.D.~Friedman, A.~Feinstein, and C.P.~Peltzer, New York, NY: Academic Press, 1968.
\item \label{rgnewton} Roger~G.~Newton, {\em Scattering Theory of Waves and Particles}, New York, NY: McGraw-Hill, 1966.
\item \label{pct} R.F.~Streater and A.S.~Wightman, {\em PCT, Spin and Statistics, and All That}, Reading, MA: W.A.~Benjamin, 1964.
\item \label{bogo} N.N.~Bogolubov, A.A.~Logunov, and I.T.~Todorov, {\em Introduction to Axiomatic Quantum Field Theory}, trans.~by Stephen Fulling and Ludmilla Popova, Reading, MA: W.A.~Benjamin, 1975.
\item \label{jost} R.~Jost, {\em The General Theory of Quantized Fields}, Providence, RI: American Mathematical Society, 1965.
\item \label{wightman-hilbert} A.S.~Wightman,``Hilbert's Sixth Problem: Mathematical Treatment of the Axioms of Physics'', {\em Mathematical Development Arising from Hilbert Problems}, ed.~by F.~E.~Browder,
{\em Symposia in Pure Mathematics 28}, Providence, RI: Amer.~Math.~Soc., 1976, p.~147. 
\item \label{wigner} T.D.~Newton and E.P.~Wigner, ``Localized States for Elementary Systems'', {\em Rev.~Modern Phys.}, Vol.~21, 1949, p.~400.
\end{enumerate}
\end{document}